# A Data-Driven, Gain-Scheduled Equivalent Circuit for the $B$–$H$ Behavior of Power Ferrites

Wilhelm Kirchgässner
*Beckhoff Automation*
wk.research@web.de

***Abstract*—An equivalent circuit for the dynamic $B$–$H$ behavior of power ferrites is identified from measured transients. It consists of classical elements of hysteresis modeling: a nonlinear inductor for the reversible curve, six viscous play branches (inductors in series with diode-bridge sliders) that form a Prandtl–Ishlinskii model, two $R$–$L$ branches for the magnetic after-effect, a mean-field feedback of $H$ as in the Jiles–Atherton model, and a measurement path. All elements are positive apart from the feedback sources, and the circuit can run in SPICE. Discretized at the sampling step, the circuit is a difference equation with 32 parameters per material, fitted by backpropagation through time. On the 15 materials of the MagNet Challenge 2, the mean sequence relative error is 7.33 ± 0.03 % against 7.77 % for the challenge-winning recurrent network with 336 parameters, and the circuit is more accurate on 12 of the 15 materials. The accuracy can be scaled up by gain scheduling, where a small network per material sets the element values at every step from the excitation, while the hysteresis memory stays in the circuit, such that the error drops to 5.25 % with 656 parameters per material. Thus, an interpretable model is demonstrated that achieves state-of-the-art accuracy.**

***Index Terms*—Hysteresis, power ferrites, play operator, Prandtl–Ishlinskii model, equivalent circuit, gain scheduling, magnetic relaxation, system identification, MagNet Challenge, interpretable machine learning**

## I. Introduction

The design of power electronics requires models of magnetic core materials that predict the field strength $H(t)$, and with it the core loss, for arbitrary excitation waveforms, frequencies and temperatures. Classical approaches are either empirical loss formulas of the Steinmetz family [1], [2] or hysteresis models such as the Preisach [3], [4], Jiles–Atherton [5], Coleman–Hodgdon [6] or Bouc–Wen [7] models, which are hard to parameterize for the rate- and temperature-dependent behavior of MnZn and NiZn ferrites at hundreds of kilohertz.

The MagNet project [8], [9], [10] and the MagNet Challenges [11], [12] turned the problem into a data-driven benchmark. The first challenge (MC1) asked for the steady-state core loss under periodic excitation where the winning entries were neural networks with a few thousand parameters per material [13]. The second challenge (MC2), built on the transient MagNetX database [14], [15], asks for the sequence-to-sequence prediction of $H$ from magnetic flux density $B$ and temperature $\vartheta$, given a short warm-up segment in which $H$ is known. The winner of the performance track, RHINO-MAG [16], is a gated recurrent unit (GRU) [17] with eight hidden states and 336 parameters per material; other strong entries use LSTMs [18] or neural operators [19] with many more parameters.

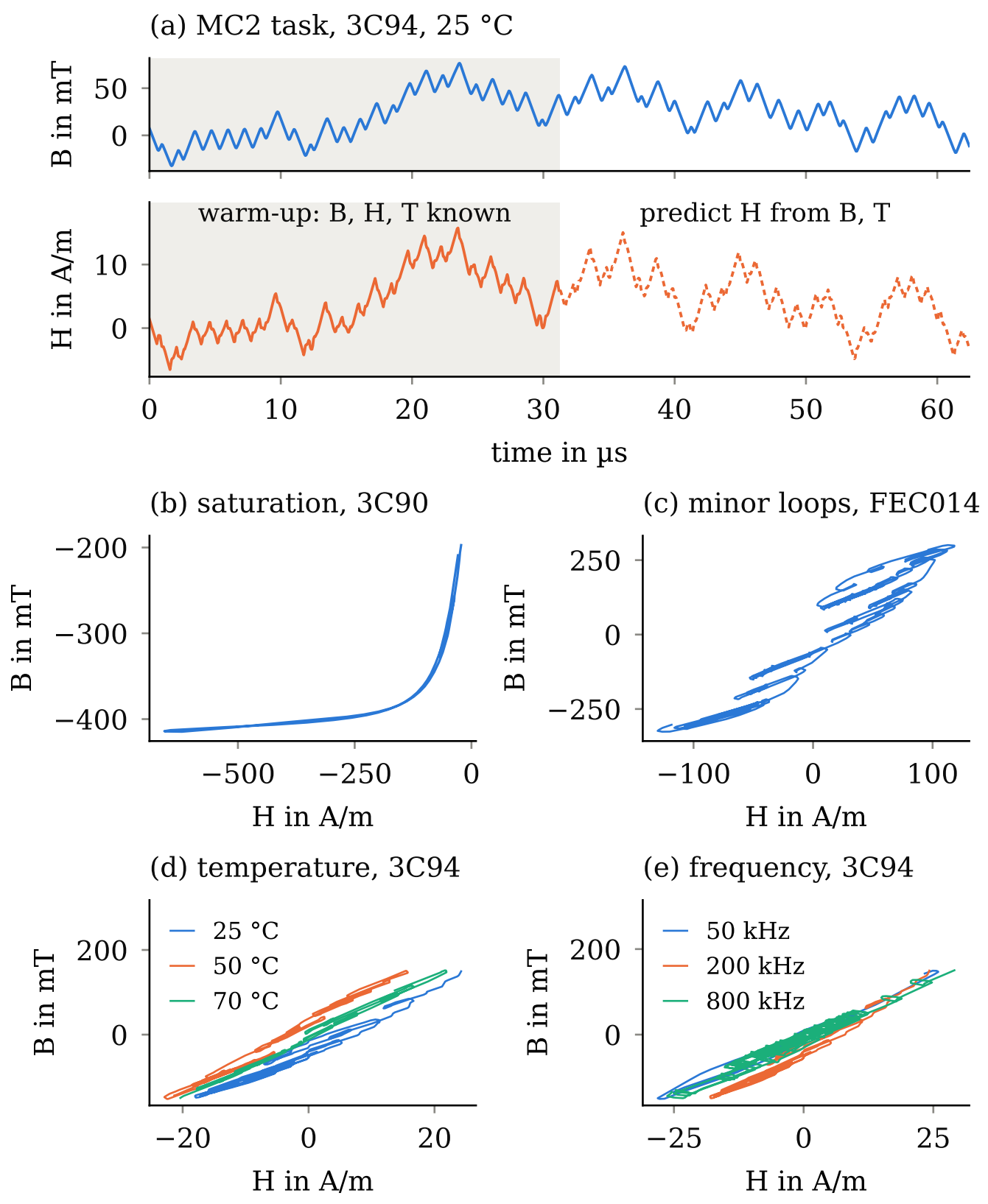

Fig. 1. The problem. (a) MagNet Challenge 2 task: in a window of 1000 samples (62.5 µs), $B$, $H$ and $\vartheta$ are known during a warm-up, and $H$ must then be predicted from $B$ and $\vartheta$ alone. The excitation is a switching ripple on a slowly wandering flux density, so the material traverses nested minor loops of all sizes. (b) Deep saturation (3C90, 50 kHz). (c) Minor loops with fast ripple on a drifting bias (FEC014, 800 kHz). (d) The same flux range at 25, 50 and 70 °C (3C94, 200 kHz) and (e) at 50, 200 and 800 kHz (3C94, 25 °C): temperature and rate impact both the slope and the width of the loops.

These models are accurate but opaque: the eight GRU states have little physical meaning, the model cannot be inspected for consistency with magnetics, and it is unclear which parts of the material behavior are learned and which are artifacts of the measurement setup. This paper asks whether the information the GRU extracts from the data can be expressed as a *governing equation* — rather than as a neural ODE [20] — with few, interpretable parameters, built from the established vocabulary of hysteresis modeling [21], [22], [23], [24], but trained on the data like a neural network.

No single element of the resulting model is new (Section II). What is new is the evidence that this minimal combination, fitted end-to-end, suffices. The contributions are:

1) **Classical elements are accurate enough.** A viscous, moving Prandtl–Ishlinskii model with 32 interpretable parameters per material is more accurate than the challenge-winning recurrent network, selected from a Pareto front, on 12 of the 15 MC2 materials with a tenth of its parameters, and an ablation shows that every element contributes (Section VII).
2) **An equivalent circuit that can be read.** The model is a circuit of positive elements apart from the feedback sources, verified in SPICE; the fit revealed a lag in the $H$ measurement, and the parameters of the material part are valid at any time step (Section IV).
3) **Gain scheduling.** Small networks that set the element values at every step, while the memory stays in the circuit, close most of the gap to the best contestants: with 64 hidden units, second of 33 by both challenge metrics (Section IV.D, Section VII.B).
4) **Extrapolation and data efficiency.** Outside its training range in frequency and temperature, the circuit degrades far less than the recurrent network (Section VII.D).

## II. Related Work

**Hysteresis operators.** Preisach models [3], [4] superpose elementary relay hysterons with a density over their switching thresholds. Prandtl–Ishlinskii (PI) models superpose play (backlash) or stop operators instead; they form the subclass of Preisach models whose density depends only on the width of the elementary loop, and their mathematical theory is well developed [21], [22], [23]. Play-type hysterons with a friction-like pinning interpretation have also been used for vector hysteresis of soft magnetic materials [25]. In mechanics the same structure is the Iwan model, a parallel bank of Jenkins elements, i.e., springs in series with Coulomb friction sliders [26]; this is the analogy behind the equivalent circuit of Section IV.A.

**Rate dependence and after-effect.** Bertotti's dynamic Preisach model lets each hysteron switch at a rate proportional to the excess of the field over its threshold [27]; the viscous play used here is the corresponding operator restricted to the PI class, with one common rate. Rate-dependent PI models with rate-dependent thresholds and densities have been developed for piezoelectric and other smart actuators [28], and Kuhnen's modified PI model adds creep operators, first-order lags, to the plays [29]. The two relaxations of the present model play the same role; physically they represent the magnetic after-effect or magnetic viscosity [30], the subject of Preisach's original paper [3]. Loss separation into hysteresis, classical eddy-current and excess contributions [31] motivates the dynamic terms.

**Mean-field feedback and circuit models.** The moving Preisach model shifts the Preisach density by a term proportional to the magnetization [32], and the Jiles–Atherton model uses the effective field $H + \alpha_{\mathrm{JA}}\mathcal{M}$ [5]; the present model, driven by $B$, feeds $H$ back into the operator inputs, the corresponding construction for a flux-driven model. Coleman–Hodgdon [6] and Bouc–Wen [7] models describe rate-independent loops with few parameters but no minor-loop memory of Preisach type. In circuit simulation, hysteresis models enter mostly as behavioral elements [33], [34]; the present model maps onto standard elements one to one.

**Gain scheduling and hypernetworks.** Gain scheduling makes the parameters of a linear controller or plant model functions of a measured scheduling variable [35], and hypernetworks generate the weights of one network with another [36]. The thermal neural network [37] combines both with a physical structure: Multilayer perceptrons (MLPs) set the conductances and power loss terms of a lumped-parameter thermal network from the operating point. The scheduled circuit in this work applies this idea to the magnetic circuit; unlike in the thermal case, the memory of the model, the hysteresis, stays in the states of the circuit. Rate-dependent PI densities [28] are an analytic form of such a schedule.

**Data-driven models.** The MagNet challenges produced accurate neural models, from convolutional networks for steady-state loss [13] to recurrent networks for transient sequences [16], [18]. Equation discovery [38] and gray-box identification [39] seek compact models instead; the approach here is gray-box, with the structure taken from the hysteresis vocabulary above and all parameters fitted jointly by gradient descent through the recurrence.

## III. Problem Statement and Data

The MC2 data [12] consist of measurements on toroidal (and, for one material, E-shaped) ferrite cores of ten materials (Ferroxcube 3C90, 3C94, 3E6, 3F4; Fair-Rite 77, 78; TDK N27, N30, N49, N87), excited with PWM-type and other transient waveforms at up to seven fundamental frequencies between 50 and 800 kHz and at 25, 50 and 70 °C. Each record holds $B$ and $H$ sampled at 16 MHz (62.5 ns per sample) over 100 periods. Five further materials (3C92, 3C95, FEC007, FEC014, T37) were used for the final evaluation of the challenge.

The task is defined on windows of 1000 samples, see Fig. 1. In the first $k$ samples both $B$ and $H$ are known (warm-up); for the remaining $1000 - k$ samples only $B$ and $\vartheta$ are given and $H$ must be predicted. Three scenarios $k \in \{100, 500, 900\}$ ("10/90", "50/50", "90/10") are evaluated. The main metric is the sequence relative error

$$\mathrm{SRE} = \frac{\sqrt{\sum_{n\geq k}\left(\hat{H}_n - H_n\right)^2}}{\sqrt{\sum_{n\geq k} H_n^2}}, \qquad (1)$$

averaged over windows and scenarios. The energy error is measured by the normalized energy relative error

$$\mathrm{NERE} = \frac{\left|\sum_{n\geq k}\left(\hat{H}_n - H_n\right)\Delta B_n\right|}{\sum_n}|H_n \Delta B_n|, \qquad (2)$$

the error of the energy $\int H\,\mathrm{d}B$ that enters the core over the predicted part, relative to the energy exchanged over the whole window. The challenge ranked its contestants by the 95th percentile of both errors, for which it normalizes the energy error by the core loss of the recording instead; Fig. 5 uses that definition. Since the official MC2 test labels of the training materials are not public, every recording is split 70/15/15 % into training, validation and test sequences, stratified by temperature, and 300 fixed test windows are drawn per frequency file. The published RHINO-MAG GRU8 models [16] are evaluated on exactly the same windows.

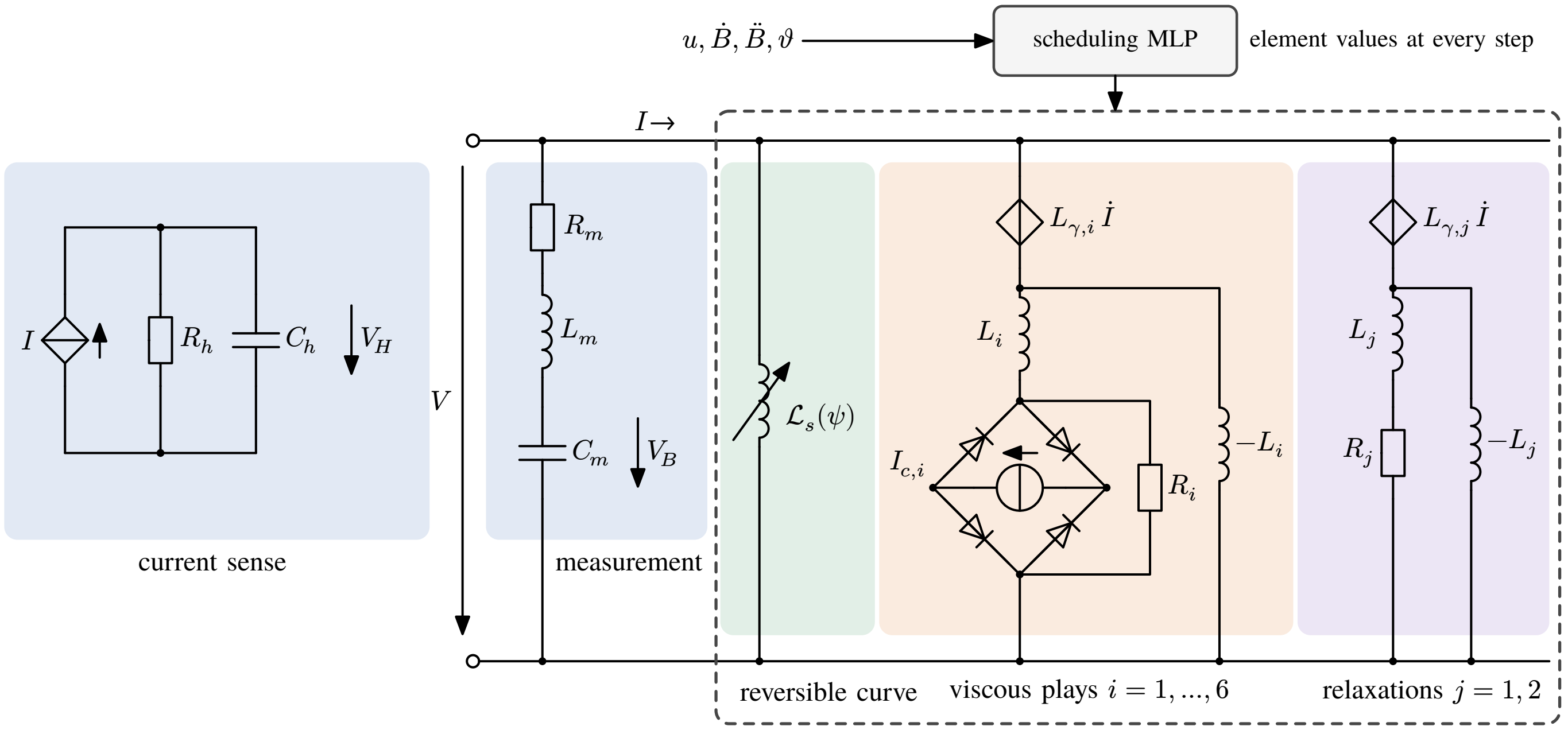

Fig. 2. Equivalent circuit of the model. The reversible curve is a flux-controlled nonlinear inductor. Each viscous play is an inductor in series with a viscous Coulomb slider: a diode bridge that carries the DC current $I_{c,i}$ (zero voltage while $|I_i| < I_{c,i}$), in parallel with $R_i$. Each relaxation is an $R$–$L$ branch. Every operator is driven through a current-controlled voltage source $L_{\gamma,i}\, dI/dt$ (the mean-field feedback). Its linear part $w_i v_i$ is the parallel inductor $-L_i$; without feedback these merge with $\mathcal{L}_s$. The measurement path reproduces the low-pass of the acquisition: the measured flux follows from $V_B$, the voltage across $C_m$; the current sense records $H$ through the lag (8) as $V_H$. Neither loads the core. Dashed: the elements whose values the scheduling MLP sets at every step in the gain-scheduled circuit (Section IV.D).

## IV. The Model

The model is presented as an equivalent circuit of the wound core in continuous time (Section IV.A). Its discretization at the sampling step is the difference equation that is fitted to the data (Section IV.B); its elements are interpreted in Section IV.C and gain-scheduled in Section IV.D. The model was found in its discrete form, by fitting candidate structures to the data; the circuit is the follow-up interpretation, not the design starting point.

### A. Equivalent Circuit

Let $u$ be the corrected flux density and $H$ the field strength of the material. With $M = 6$ viscous plays and $J = 2$ relaxations, the model reads

$$u = B + \tau_1 \dot{B} + \tau_2 \ddot{B}, \tag{3}$$

$$v_i = u + \gamma_i H, \tag{4}$$

$$\dot{p}_i = \kappa_\alpha\, \delta_{r_i}(v_i - p_i), \quad i \le M, \tag{5}$$

$$\dot{z}_j = \kappa_j \big(v_{M+j} - z_j\big), \quad j \le J, \tag{6}$$

$$H = h_s(u) + \sum_{i=1}^{M} w_i p_i + \sum_{j=1}^{J} c_j z_j + H_{\text{off}}, \tag{7}$$

$$\tau_h \dot{H}^{\text{m}} = H - H^{\text{m}}, \tag{8}$$

with the dead-zone function $\delta_r(x) = \operatorname{sign}(x)\max(|x| - r, 0)$ and the reversible curve

$$h_s(u) = a_1 u + a_3 u^3 + H_{\text{kn}}\big[e^{(u-B_s)/b} - e^{(-u-B_s)/b}\big], \tag{9}$$

$H_{\text{kn}} = 10$ A/m. A play state $p_i$ follows its drive $v_i$ only once the drive leaves the band of width $r_i$ around it, and then with the rate $\kappa_\alpha$; a relaxation state $z_j$ follows its drive as a first-order lag. $H^{\text{m}}$ is the field strength as the instrument records it.

For a core with $N$ turns, cross-section $A_e$ and path length $l_e$ the flux linkage is $\psi = N A_e u$ and the winding current $I = l_e H/N$. Each additive term of $H$ is then a current branch in parallel across the winding (Fig. 2): the reversible curve is the nonlinear inductor $I = (l_e/N)\, h_s(\psi/(N A_e))$, a term $w_i p_i$ is the play inductor $L_i = N^2 A_e/(l_e\, |w_i|)$ with the slider current $I_{c,i} = l_e\, |w_i|\, r_i/N$ and the slider resistor $R_i = \kappa_\alpha L_i$, a relaxation is the branch $L_j = N^2 A_e/\big(l_e\, |c_j|\big)$, $R_j = \kappa_j L_j$, and the feedback source is $L_{\gamma,i}\, dI/dt$ with $L_{\gamma,i} = N^2 A_e \gamma_i / l_e$. The measurement path has $R_m C_m = \tau_1$, $L_m C_m = \tau_2$ and $R_h C_h = \tau_h$. All values are linear or constant in $\vartheta$, as the model's coefficients.

The play branches are the electrical form of the Jenkins element of mechanics, a spring in series with a Coulomb friction slider; a parallel bank of them is the Iwan model, the mechanical form of the Prandtl–Ishlinskii operator. While its current stays below $I_{c,i}$ the slider does not move and the branch is a plain inductor. Above it, the excess current flows through $R_i$, which sets the viscous drag of the play. This matches the picture of domain walls that are pinned until the field exceeds a threshold and then move against eddy-current damping.

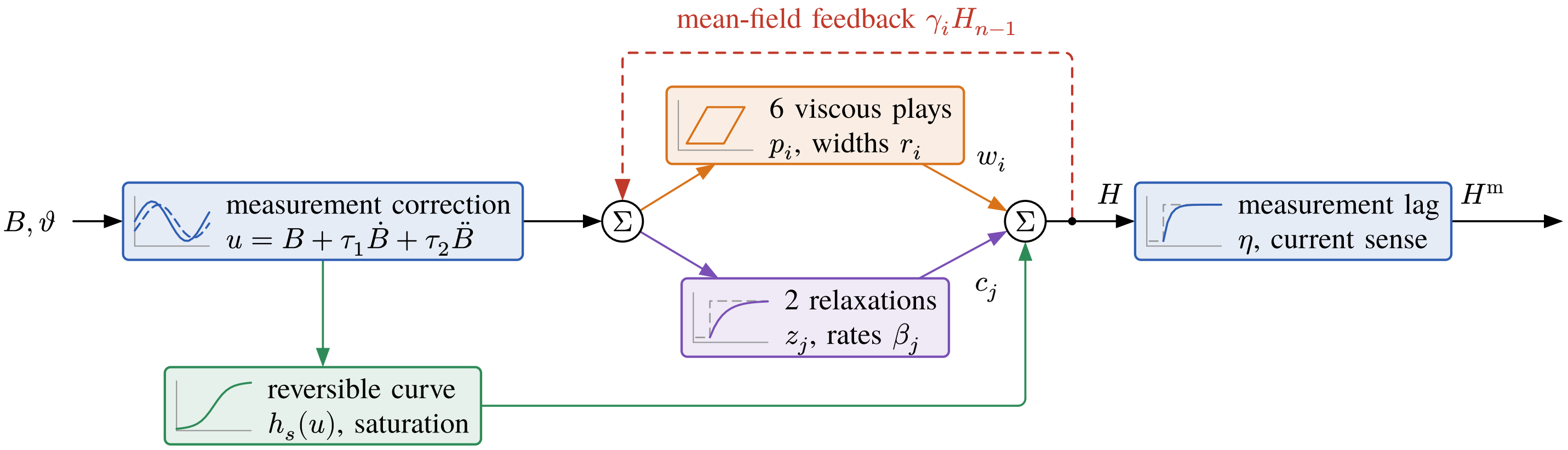

Fig. 3. Structure of the model. The play and relaxation states are driven by the corrected flux density plus a feedback of the field strength; $H$ is the sum of a reversible curve and weighted states, and the instrument records it through a first-order lag. During warm-up an observer (not shown) corrects the states with the measured $H$.

The model is trained with the constraints $w_i \leq 0$ and $c_j \leq 0$ (Section IV.B), so that every play and relaxation branch consists of positive elements; an unconstrained fit already gives $w_i < 0$ for all materials, but mostly $c_j > 0$, i.e., negative $R$–$L$ branches. The bound $\beta_j \leq 1/2$ keeps the relaxation time constants above 1.4 samples, below which the discrete model no longer approximates its continuous reading. The parallel inductors $-L_i$, $-L_j$ combine with the reversible inductor to a linear inductance that is positive for all materials at 25 and 70 °C except 3E6 at 70 °C; apart from that case, the circuit without the feedback sources is passive. The feedback sources are active, as is the effective field of Jiles–Atherton theory. The measurement path does not load the core.

*B. Discretization and Step Size*

The circuit is fitted at the sampling step $\Delta t = 62.5$ ns of the data. Over one step, the drive of every operator is held constant (zero-order hold) and the feedback uses the field of the previous sample, which breaks the algebraic loop between (4) and (7). For a constant drive, (5), (6) and (8) have exact solutions over one step, so that the rates become $\alpha = 1 - e^{-\kappa_\alpha \Delta t}$, $\beta_j = 1 - e^{-\kappa_j \Delta t}$ and the lag pole $\eta = e^{-\Delta t/\tau_h}$; the derivatives in (3) become central differences in units of samples. Let $n$ be the sample index and $\tilde{\vartheta} = \frac{\vartheta - 50°\mathrm{C}}{25°\mathrm{C}}$ the normalized temperature. Every coefficient written as $\varphi(\vartheta)$ below is linear in $\tilde{\vartheta}$, $\varphi(\vartheta) = \varphi_0 + \varphi_1 \tilde{\vartheta}$. The discrete model (Fig. 3) reads

$$u_n = B_n + \tau_1(\vartheta)\dot{B}_n + \tau_2(\vartheta)\ddot{B}_n, \tag{10}$$

$$v_{i,n} = u_n + \gamma_i(\vartheta) H_{n-1}, \tag{11}$$

$$p_{i,n} = p_{i,n-1} + \alpha_i(\vartheta)\, \delta_{r_i(\vartheta)}\big(v_{i,n} - p_{i,n-1}\big), \tag{12}$$

$$H_n = h_s(u_n; \vartheta) + \sum_i w_i(\vartheta)\, p_{i,n} + H_{\mathrm{off}}, \tag{13}$$

$$H_n^{\mathrm{m}} = \eta\, H_{n-1}^{\mathrm{m}} + (1-\eta) H_n, \tag{14}$$

for $i = 1, \ldots, M + J$, where the relaxations are written as operators of zero width, $p_{M+j} = z_j$, $r_{M+j} = 0$, $w_{M+j} = c_j$, with their own rates $\alpha_{M+j} = \beta_j$; the plays share the rate $\alpha$. $H^{\mathrm{m}}$ is the quantity that is predicted and, during the warm-up, measured. The feedback (11) uses the field $H$ of the material, not the recorded $H^{\mathrm{m}}$; during the warm-up, $H$ is the de-lagged measurement. Only the offset $H_{\mathrm{off}}$ is not a parameter; it is part of the initial state.

The operators are not parameterized individually. The play widths are log-equidistant between $r_1$ and $r_M$ with a common temperature factor,

$$\ln r_i(\vartheta) = \ln r_1 + \frac{i-1}{M-1}(\ln r_M - \ln r_1) + \rho\tilde{\vartheta}. \tag{15}$$

The play weights follow a density that is linear in the standardized logarithm $\chi_i$ of the width (a linear function of the play index from $-1.34$ to $1.34$), $w_i(\vartheta) = d_0(\vartheta) + d_1(\vartheta)\chi_i$, and the feedback gains are $\gamma_i = \gamma(\vartheta) + g_p \chi_i$ for the plays and $\gamma(\vartheta) + g_j$ for the relaxations. Again, the circuit is constrained to consist of positive elements apart from the feedback (Section IV.A): $w_i \leq 0$, $c_j \leq 0$ and $\beta_j \leq 1/2$. With the pole $\eta$ of the measurement lag, this gives 32 parameters per material (Table I).

TABLE I

PARAMETERS OF THE MODEL PER MATERIAL. FIVE BELONG TO THE MEASUREMENT CHAIN, 27 TO THE MATERIAL. THE OBSERVER HAS ITS OWN PARAMETERS, WHICH ARE NOT COUNTED (SECTION V).

| **Term** | **#** | **Parameters** |
|---|---|---|
| measurement | 5 | $\tau_1, \tau_2$ linear in $\tilde{\vartheta}$; lag pole $\eta$ |
| reversible | 7 | $a_1, a_3, B_s$ linear in $\tilde{\vartheta}$; knee width $b$ |
| play widths | 3 | $r_1, r_6$ (log-equidistant); temperature slope $\rho$ |
| play rate | 2 | $\alpha = \sigma\big(\alpha_0 + \alpha_1\tilde{\vartheta}\big)$ |
| play density | 4 | $d_0, d_1$ linear in $\tilde{\vartheta}$ |
| relaxations | 6 | $\beta_1, \beta_2$; $c_1, c_2$ linear in $\tilde{\vartheta}$ |
| feedback | 5 | $\gamma$ linear in $\tilde{\vartheta}$; $g_p$; $g_1, g_2$ |
| **total** | **32** | |

**Step size.** Since the material part of the model is the discretization of (4)–(7), it can be evaluated at any step $\Delta t$ by $\alpha \to 1-(1-\alpha)^{\Delta t/62.5\text{ ns}}$. Run open-loop on test windows of all 15 materials with 8 sub-steps per sample, the model changes by a median of 0.07 % (RMS, relative) against the native step, most for FEC014 (median 0.6 %), which combines a fast play rate with a relaxation at its bound. Applied on a finer grid, the measurement correction instead turns the corners of the PWM waveforms into ever narrower spikes (median deviation 3.8 % at 8-fold upsampling for 3C94): it is a property of the 16 MHz acquisition, so $u$ is computed on the 16 MHz grid and then resampled, and the lag is omitted when the true field is wanted.

**SPICE.** In ngspice 42, with the slider written as the dead-zone resistor $v = R_i\, d_{I_c}(i)$ that the diode bridge implements, the circuit driven with the corrected flux of 60 test windows reproduces the continuous-time model with a median deviation of 0.01 % (maximum 0.38 %): the circuit is an exact realization of the model.

### C. Interpretation

**Measurement correction (10).** The term inverts, to second order, a low-pass $B_{\text{meas}}(s) = B(s)/(1+\tau_1 s+\tau_2 s^2+\ldots)$ between the true and the recorded flux density (or a relative delay of the voltage and current channels plus the winding's parasitic dynamics), a property of the specimen and its winding, not of the material.

**Measurement lag (14).** The recorded field strength follows the field of the model through a first-order lag with pole $\eta$. It was not designed in but found in the fitted model. Without the lag and without the passivity constraints, the model drives both relaxations of 3C92 and FEC014 to $\beta_j = 1$ with $c_j > 0$, in every one of ten training seeds. Such a relaxation follows its drive instantly, $z_j = u + \gamma_j H_{n-1}$, so that the two together add $\sum_j c_j \gamma_j H_{n-1}$ to $H_n$: an autoregression of the output with coefficient 0.59 (3C92), 0.51 (FEC014) and 0.1 (N87), and close to zero for all other materials. The autoregression is a first-order lag of $H$, which a passive model cannot form; with the lag as an explicit element, the passive model is as accurate as the unconstrained one. The fitted poles are at or close to the bound $\eta \le 0.6$ (about 0.1 µs) for N87, 3C92, 3C95, FEC014 and N27, and below 0.4 for the rest. A lag in the current channel of the acquisition, e.g., of the current probe or shunt, is the natural reading. During the warm-up, the measured $H^{\text{m}}$ is de-lagged by the exact inverse of (14), which amplifies noise by $(1+\eta)/(1-\eta)$; hence the bound on $\eta$.

**Reversible curve (9).** A linear term, a cubic term and a smooth exponential knee at $\pm B_s(\vartheta)$ describe the anhysteretic part; the knee is needed for the low-frequency recordings that reach saturation.

**Viscous plays.** For $\alpha = 1$ each state is the classical play operator, and $\sum w_i p_i$ is a Prandtl–Ishlinskii model, i.e., a Preisach model with a one-dimensional density [21], [22]; $-w_i$ is the discretized density, and for all materials wide plays carry more weight than narrow ones. With $\alpha < 1$ the operator follows its input with a first-order lag once the dead zone is exceeded, a discrete form of the dry-friction pinning picture [25] with a viscous term, which makes the loops widen with $\dot{B}$ as observed for ferrites [31], [34]. The fitted time constants are 4–28 samples (0.25–1.8 µs); only on 3F4 are the plays rate-independent.

**Relaxations and feedback.** With $c_j \le 0$, a step of the drive raises $H$ at once and lets it decay toward the static value, as an eddy-current field would: magnetic after-effect, the time-domain counterpart of the relaxation term of the i²GSE [2]. The slow relaxation has time constants of 2.5–35 µs for most materials, the fast one at most ten samples on nine materials. The mean feedback gain of the plays is 0.2–12 mT per A/m; since $H$ reaches several hundred A/m, $\gamma H$ is comparable to the flux amplitude, as the effective field of the Jiles–Atherton model [5] written in the $B$-driven direction.

**Specimen and material.** The fitted parameters of all materials and their transfer to other specimens are given in the supplementary material[1]. The raw MC1 data [11] contain the ten training materials measured with the same instruments on other cores, mostly with other numbers of turns. Refitting only the five measurement parameters of the MC2 model on these recordings recovers two thirds of the gap between no refit (10.8 % SRE) and a full refit (7.1 %), the 14 amplitude parameters together with them 84 %, and the remaining 13 dynamic parameters add 0.6 points. The parameters thus fall into three tiers: the measurement chain differs most between setups, the amplitudes follow the specimen, and the dynamics carry over almost unchanged.

### D. Gain Scheduling

The element values of the circuit depend on temperature only. Its remaining error is largely structural: the constraints that make it readable, one rate for all plays, a density linear in $\ln r$, two relaxations and element values that are constant in time, also limit how closely it can follow the data. Gain scheduling relaxes these limits but keeps the structure: the element values become functions of the excitation, as in gain-scheduled control [35] and in the thermal neural network [37].

Per material, an MLP $f$ with two hidden layers of $h$ tanh units maps the scheduling variables

$$\boldsymbol{x}_n = \Big(u_n/u_0, \operatorname{asinh}\big(\dot{B}_n/b_1\big), \operatorname{asinh}\big(\ddot{B}_n/b_2\big), \tilde{\vartheta}\Big), \quad (16)$$

with $u_0 = 0.2$ T, $b_1 = 1$ mT and $b_2 = 0.1$ mT (derivatives per sample), to 16 shifts

$$\boldsymbol{z}_n = \frac{1}{2}[f(\boldsymbol{x}_n) + f(\boldsymbol{S}\boldsymbol{x}_n)], \quad (17)$$

where $\boldsymbol{S} = \operatorname{diag}(-1,-1,-1,1)$ reverses the excitation, so that the model stays odd, $H(-B) = -H(B)$. In the circuit, every coefficient $\varphi$ is computed from an unconstrained raw value $\varphi^{\text{raw}}$ through a fixed map $g_\varphi$ that enforces its constraint, e.g., a logistic function for $\alpha$ and $\beta_j$, $-\operatorname{softplus}(-\,\cdot)$ for $w_i$ and $-\operatorname{softplus}$ for $c_j$. The schedule adds the shift to the raw value,

$$\varphi_n = g_\varphi\Big(\varphi^{\text{raw}}(\vartheta) + z_{\varphi,n}\Big). \quad (18)$$

[1] https://github.com/wkirgsn/ferrite-bh-circuit

The 16 shifts act on five output-side coefficients, $a_1$, $a_3$, $B_s$, the knee width $b$ and a gain $e^z$ of the whole hysteretic part $\sum_i w_i p_i$, and on eleven dynamic ones: the play rate $\alpha$, the density coefficients $d_0$ and $d_1$, the relaxation rates $\beta_j$ and weights $c_j$, and two shifts each of the play widths and the feedback gains, a common one and one that tilts them along the play index,

$$\begin{aligned} \ln r_{i,n} &= \ln r_i(\vartheta) + z_{r,n} + z'_{r,n}\,\chi_i, \\ \gamma_{i,n} &= \gamma_i(\vartheta) + z_{\gamma,n} + z'_{\gamma,n}\,\chi_i, \end{aligned} \tag{19}$$

with $\chi_i = 0$ for the feedback of the relaxations. The last layer of $f$ is initialized to zero, so that training starts at the classical circuit, the special case $\boldsymbol{z} = 0$. With $h = 16$ hidden units the network adds 624 parameters, 656 per material in total; with $h = 64$ it adds 5520, 5552 in total. The 32 parameters of the classical circuit are trained jointly with the network; they become the operating point of the scheduled circuit and need not equal the classical fit.

The scheduling variables depend on $B$ and $\vartheta$ only, so the shifts of a whole window are computed before the recurrence, and the network has no state of its own: the memory of the model remains in the play and relaxation states. In the circuit (dashed in Fig. 2), the nonlinear inductor, the play and relaxation branches and the feedback gains become time-varying elements. Because every value passes through the same map as in the classical circuit, every element is positive at every step. A time-varying inductor is nevertheless not passive: changing its value at a fixed flux changes its stored energy, which is quantified in Section VII.B.

## V. State Estimation During Warm-Up

In every test window the initial state — the magnetization history encoded in the play operators — is unknown. It is estimated from the warm-up by an estimator that is kept separate from the model; its parameters are not counted, since they describe the initial state and not the material; in circuit simulation the model starts from a known (e.g., demagnetized) state. All quantities below use the de-lagged $H$.

**Prior.** A play state lies within its dead zone around the drive, $|u - p_i| \le r_i$, but where depends on the history before the window. The prior places each play within its band as a smooth function of the level $u_0$ and the direction of the excitation at the first sample,

$$p_i = u_0 - r_i \tanh\bigl(\xi_{2,i} u_0 / r_i + \xi_i \operatorname{sign}(\dot{u}_0)\bigr), \tag{20}$$

with two learned coefficients $\xi_i$ and $\xi_{2,i}$ per play and material. A large $\xi_i$ puts the play at the trailing edge of its band, as on a branch along which it is dragged by the drive; $\xi_{2,i} > 0$ puts it below the drive at a high level, as at the top of a loop reached from below. The fit uses the direction mainly for the narrowest play. The relaxations start at $u_0$.

**Mismatch distribution and observer.** The prior leaves a mismatch $e_0 = H_0 - \hat{H}_0$. Absorbing it in the offset $H_{\text{off}}$ alone is correct only until the first reversal of $B$, after which the plays move and the wrong offset becomes visible as a jump; it is therefore distributed over all states with learned gains $K_0$, $p_i \leftarrow p_i + K_{0,i} e_0$, $H_{\text{off}} = H_{\text{pr}} + K_{0,\text{off}} e_0$. During the warm-up, the innovation $e_n = H_n - \hat{H}_n$ corrects every state with learned constant gains, as in a Luenberger observer [40].

**Amplitude estimate.** The amplitude of $H$ varies from window to window beyond what $B$ and $\vartheta$ explain: it is set by the excitation history before the window. Two factors $q$ and $q_h$ for the reversible curve without the knee and for the hysteretic part are fitted to the detrended mismatch of an open-loop pass over the warm-up by a Bayesian ridge regression whose prior is worth $\nu_0 = e^{\lambda_0}(k/300)^{\lambda_1}$ samples of the window's mean square field, clamped to $|q| \le 0.5$, and the main pass scales both parts accordingly. Long warm-ups thus determine the amplitudes from the data, short ones fall back to the model's own. During training, a penalty $0.5(q^2 + q_h^2)/2$ makes the model carry the mean amplitude of a material.

Since the whole warm-up is given at once, the estimator need not be causal; but for the state at the end of the warm-up, which the prediction needs, the causal observer already uses all the warm-up data, and choosing the initial state among candidate reversal histories by the open-loop error was worse at every warm-up length.

In total, the estimator learns 33 parameters per material, which are not part of the 32 of the model: the 12 coefficients $\xi_i, \xi_{2,i}$ of the play prior, the offset prior $H_{\text{pr}}$, the nine gains $K_0$ that distribute the initial mismatch, the nine observer gains and the two coefficients $\lambda_0, \lambda_1$ of the amplitude prior.

## VI. Implementation and Training

The model is a nonlinear recursion over up to $10^3$ samples per window and cannot be parallelized over time. A straightforward PyTorch [41] loop is limited by kernel launch overhead (about 0.4 s per optimizer step regardless of batch size). The forward pass and its hand-derived adjoint were therefore written as a single fused Triton [42] kernel in which each GPU thread processes one window and keeps all $M + J$ states in registers; the backward pass stores the states at every step and recomputes the step-local Jacobians. On an AMD RX 9070 XT this increases the throughput by a factor of about 400, which makes gradient-based fitting to $10^9$ samples practical.

Parameters are fitted with Adam [43] and a one-cycle learning-rate schedule [44] (peak $2 \cdot 10^{-2}$, 20 000 steps). Each step draws 768 random windows of 1000 samples per material, weighted by the number of sequences per frequency file, and a random warm-up length $k \in \{100, 500, 900\}$ as in the evaluation. The loss is the mean of SRE + NERE of the predicted part, summed over materials. Initial values of $a_1$ and $w_i$ are scaled by the least-squares slope of $H$ over $B$ of each material. Several model variants are trained simultaneously in one kernel launch. One training run of all 15 materials takes about one hour.

**Scheduled circuit.** The kernel takes the operator coefficients per step, and its adjoint returns their gradients per step, through which the network is trained. Training starts from the fitted classical circuit, with the same warm-up estimator and loss, 512 windows per material and step, 20 000 steps, and peak learning rates of $10^{-3}$ for the network and $2 \cdot 10^{-3}$ for

TABLE II

Test results in % of the published RHINO-MAG GRU8 (336 parameters) and of the classical circuit (32 parameters) on identical windows: mean and 95th percentile of the sequence relative error (SRE) and of the normalized energy error (NERE), each per scenario and averaged over the three scenarios; the classical circuit's mean SRE per scenario. Bold: better than GRU8. The final-evaluation materials 3C92, 3C95, FEC007, FEC014 and T37 are evaluated on the official MC2 test set.

| | SRE | | SRE, 95th perc. | | NERE | | NERE, 95th perc. | | SRE per scenario | | |
|---|---|---|---|---|---|---|---|---|---|---|---|
| **Material** | GRU8 | classical | GRU8 | classical | GRU8 | classical | GRU8 | classical | 10/90 | 50/50 | 90/10 |
| 3C90 | 7.91 | **7.68** | 22.54 | **21.05** | 0.72 | **0.68** | 2.63 | **2.03** | 9.16 | **7.50** | **6.37** |
| 3C94 | 7.47 | **6.45** | 21.89 | **16.90** | 0.64 | **0.55** | 2.28 | **1.77** | 8.20 | **6.00** | **5.15** |
| 3E6 | 10.22 | **10.08** | 24.46 | **23.19** | 1.10 | 1.22 | 3.24 | 3.71 | 11.07 | **9.70** | 9.47 |
| 3F4 | 4.83 | **4.61** | 14.52 | **12.27** | 0.45 | **0.42** | 1.68 | **1.56** | 7.06 | **3.85** | **2.93** |
| 77 | 8.44 | **7.46** | 24.80 | **19.94** | 0.79 | **0.71** | 2.73 | **2.31** | **9.05** | **6.95** | **6.38** |
| 78 | 7.15 | **6.25** | 21.18 | **16.53** | 0.67 | **0.59** | 2.41 | **1.85** | **7.88** | **5.86** | **5.01** |
| N27 | 8.35 | **7.42** | 26.68 | **20.10** | 0.75 | **0.64** | 2.73 | **1.96** | 9.37 | **6.85** | **6.05** |
| N30 | 8.24 | **7.63** | 21.16 | **19.03** | 0.78 | 0.79 | 2.39 | **2.34** | 8.92 | **7.08** | **6.89** |
| N49 | 8.39 | **7.14** | 22.31 | **20.32** | 0.63 | 0.63 | 1.81 | 2.31 | **9.25** | **6.84** | **5.33** |
| N87 | 6.86 | 7.14 | 17.94 | 18.09 | 0.70 | 0.79 | 2.26 | 2.50 | 9.62 | **6.33** | **5.46** |
| 3C92 | 10.97 | 12.34 | 23.37 | 31.90 | 1.27 | 2.45 | 3.83 | 10.71 | 14.04 | 11.70 | 11.26 |
| 3C95 | 6.51 | **4.95** | 17.58 | **13.81** | 0.51 | 0.71 | 1.35 | 1.82 | **6.69** | **3.97** | **4.18** |
| FEC007 | 8.05 | **6.69** | 22.65 | **16.56** | 0.73 | **0.58** | 2.67 | **1.73** | **7.70** | **6.20** | **6.16** |
| FEC014 | 6.04 | 7.85 | 16.25 | 18.62 | 0.58 | 0.94 | 2.01 | 3.30 | 10.35 | 8.02 | 5.19 |
| T37 | 7.16 | **5.89** | 18.50 | **13.40** | 0.69 | **0.64** | 2.33 | **2.10** | **6.34** | **5.50** | **5.82** |
| **mean** | 7.77 | **7.31** | 21.06 | **18.78** | 0.73 | 0.82 | 2.42 | 2.80 | 8.98 | **6.82** | **6.11** |
| **better** | | 12/15 | | 12/15 | | 8/15 | | 9/15 | | | |

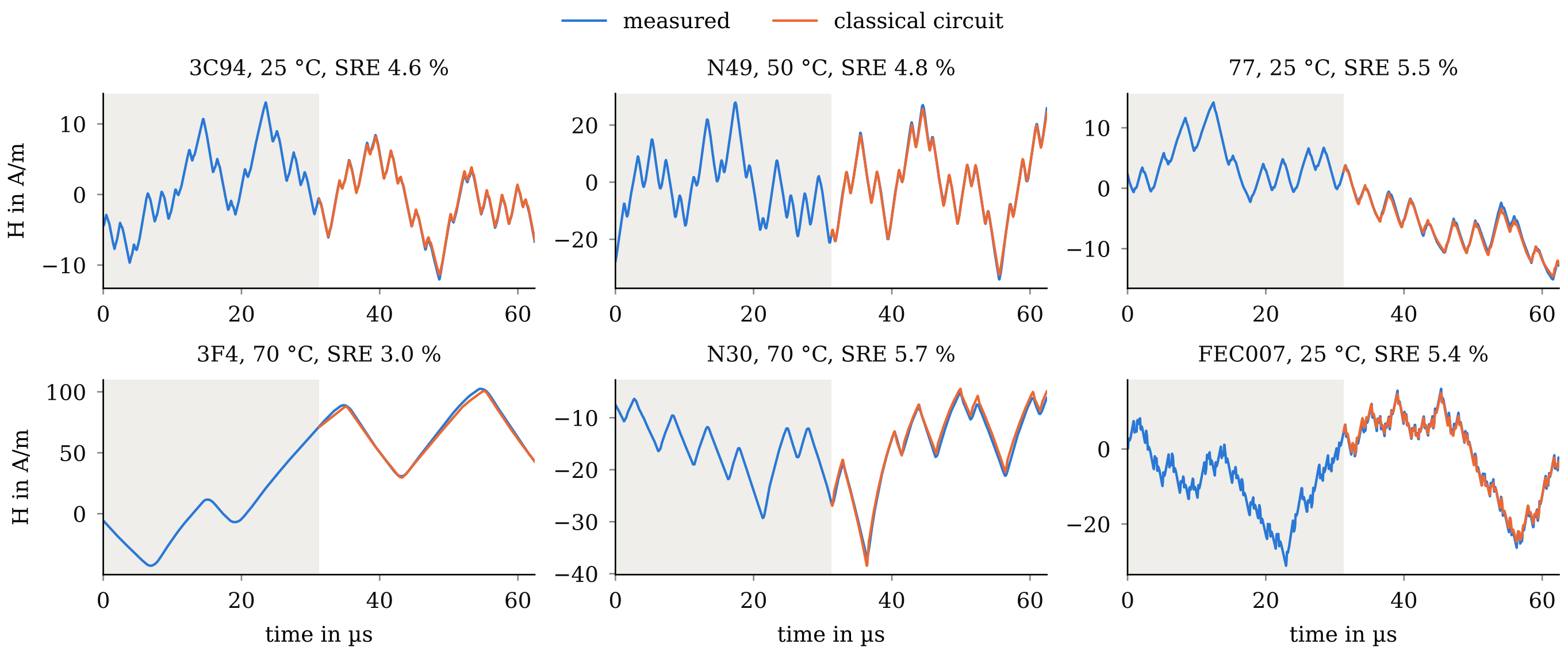

Fig. 4. Representative test windows (median error of the model in the 50/50 scenario). Shaded: warm-up with known $H$; afterwards the model predicts $H$ from $B$ and $\vartheta$ alone. FEC007 belongs to the official MC2 test set.

the 32 circuit parameters; one run of all 15 materials takes about one hour with 16 hidden units and two hours with 64.

# VII. Results

## A. Accuracy

Table II compares the classical circuit (32 parameters, passive, with measurement lag) with the published GRU8

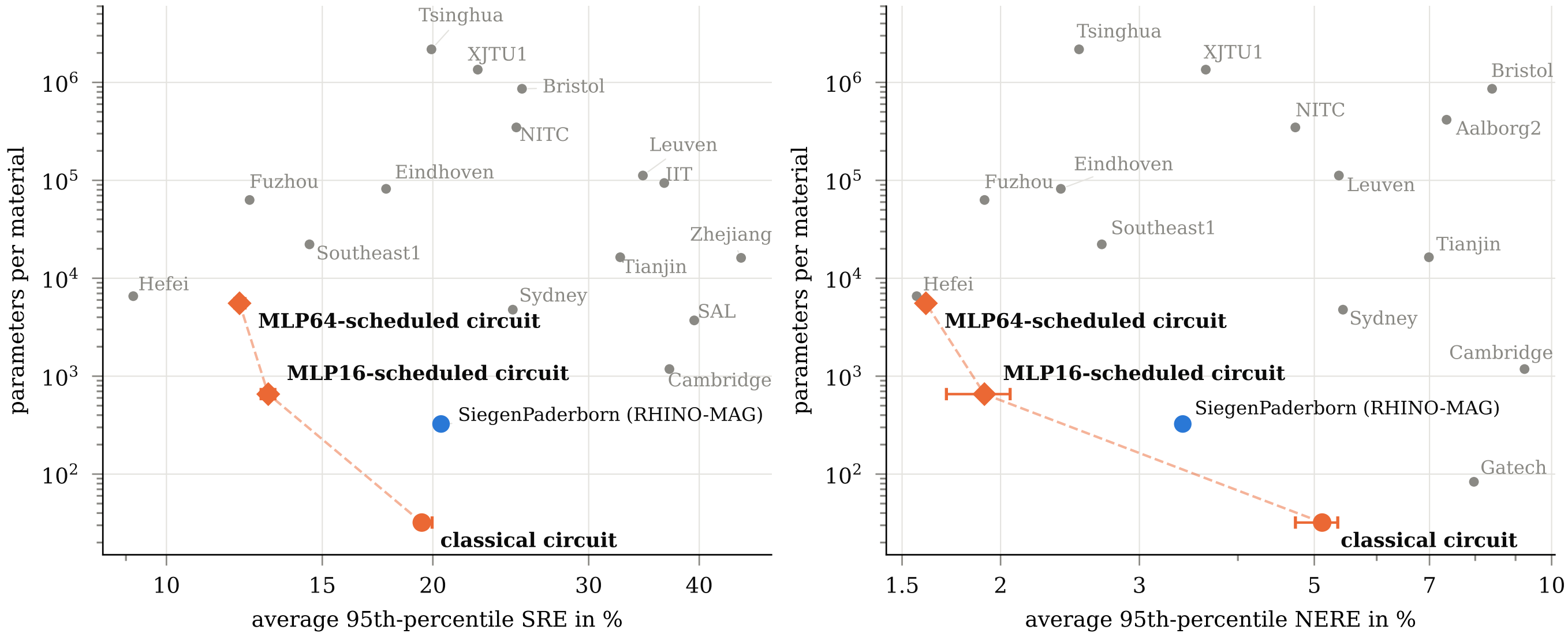

Fig. 5. The classical and the gain-scheduled circuits among the contestants of the MagNet Challenge 2: parameters per material over the average 95th percentile of the SRE (left) and of the energy error (right) on the official test set of the five final-evaluation materials; contestants from the official results, the better half by each metric shown (energy error: up to 10 %). Circuits: mean and range over the training seeds (classical: ten, scheduled: five); scheduled circuits with 16 or 64 hidden units.

TABLE III
Classical and gain-scheduled circuits (scheduled: reversible curve, hysteresis gain and operator dynamics, trained jointly with the 32 circuit parameters, 20 000 steps): hidden units $h$, parameters per material (without the 33 of the warm-up estimator, Section V, which all three circuits share), mean test SRE and NERE in % over the 15 materials, and the official average 95th-percentile SRE and energy error in % on the five final-evaluation materials. Mean ± standard deviation over the number of training seeds in brackets.

| Circuit | $h$ | Param. | Mean test | | Official p95 | |
|---|---|---|---|---|---|---|
| | | | SRE | NERE | SRE | NERE |
| classical (10) | – | 32 | 7.33 ± 0.03 | 0.82 ± 0.01 | 19.43 ± 0.22 | 5.12 ± 0.19 |
| MLP16-scheduled (5) | 16 | 656 | 5.25 ± 0.04 | 0.45 ± 0.00 | 13.03 ± 0.17 | 1.91 ± 0.14 |
| MLP64-scheduled (5) | 64 | 5552 | 4.80 ± 0.03 | 0.39 ± 0.00 | 12.09 ± 0.14 | 1.61 ± 0.01 |

models of RHINO-MAG on identical test windows, with one configuration for all 15 materials; Fig. 4 shows representative predictions. Over ten training seeds the mean SRE is 7.33 ± 0.03 % against 7.77 % for the GRU (7.31 % for the run in Table II), and the model is more accurate on 12 of the 15 materials. Its advantage is largest in the 50/50 and 90/10 scenarios (6.85 % and 6.11 % against 8.14 % and 6.59 %), where the warm-up determines the amplitude estimate well; with a 100-sample warm-up it is behind (9.04 % vs 8.59 %). The 95th percentile of the SRE is lower than the GRU's on the same 12 materials (18.78 % vs 21.06 %). The energy error is the weaker side: the mean NERE is 0.82 % against 0.73 %, better on 8 of the 15 materials, with most of the difference on 3C92 (95th percentile 10.7 % vs 3.8 %).

With 32 parameters per material, the classical circuit is smaller than every submitted model, the smallest of which had 83 (Fig. 5), and its average 95th-percentile SRE on the official test set is on par with that of the challenge winner RHINO-MAG (19.4 % vs 20.4 %). Gain-scheduled with 16 hidden units (656 parameters), it would rank third by the SRE (13.0 %) and second by the energy error (1.91 %, level with Fuzhou); with 64 hidden units (5552 parameters), second by both (12.1 % and 1.61 %; five seeds each). Only Hefei, with about 6600 parameters per material, is more accurate by both metrics.

The GRU remains better on three materials, for different reasons. On N87 the model is behind only with the short warm-up (9.6 % vs 7.8 % at $k = 100$) and at 50 and 80 kHz, where the excitation reaches deep saturation and the single exponential knee is a rough description. On 3C92 it is behind at every warm-up length; with only 15 training sequences per frequency, the saturation knee is poorly determined at 70 °C. On FEC014 it is behind even at $k = 900$ (5.2 % vs 4.1 %), a deficit of the model structure rather than of state estimation; the lag pole of FEC014 sits at its bound.

### B. *Gain Scheduling*

Table III compares the scheduled circuits with the classical one. With 16 hidden units, the mean test SRE is 5.25 ± 0.04 % (five seeds), 28 % below the classical circuit and 32 % below the GRU, and the mean NERE drops from 0.82 % to 0.45 %; with 64 hidden units, the SRE is 4.80 ± 0.03 % and the NERE 0.39 % (five seeds). On the official test set, the average 95th-percentile SRE falls from 19.4 % to 13.0 ± 0.2 % and 12.1 ± 0.1 %, and the energy error from 5.1 % to 1.9 ± 0.1 % and 1.61 ± 0.01 %.

TABLE IV

Ablation of the classical circuit: each row removes one classical element, all variants trained jointly with the same data and seed. Mean test SRE in % over the 15 materials, overall and per scenario; $\Delta$: change against the classical circuit; W: materials (of 15) on which the variant is more accurate than GRU8.

| Model | #par | SRE | $\Delta$ | 10/90 | 50/50 | 90/10 | W |
|---|---|---|---|---|---|---|---|
| classical circuit | 32 | 7.30 | | 8.98 | 6.82 | 6.11 | 12 |
| without saturation knee | 29 | 7.91 | +0.61 | 9.83 | 7.43 | 6.48 | 8 |
| without cubic term | 30 | 7.60 | +0.29 | 9.32 | 7.17 | 6.30 | 10 |
| rate-independent plays ($\alpha = 1$) | 30 | 10.99 | +3.68 | 13.23 | 10.60 | 9.14 | 10 |
| without relaxations | 24 | 7.68 | +0.38 | 9.37 | 7.26 | 6.42 | 11 |
| without mean-field feedback | 27 | 8.09 | +0.79 | 10.06 | 7.59 | 6.62 | 11 |
| without measurement correction of $B$ | 28 | 12.62 | +5.32 | 13.10 | 12.09 | 12.68 | 0 |
| without lag of $H$ | 31 | 7.46 | +0.16 | 9.18 | 6.95 | 6.25 | 12 |

The schedule acts mainly at the reversals of $B$: on 3C94 and N87, for example, the play rate and the widths of the narrow plays change abruptly at every reversal and relax along the following branch. On 3C94 the play weights also rise with $|B|$ by up to 30 %, i.e., the hysteresis density depends on the amplitude. Jointly trained, the operating point moves away from the classical fit: the mean play rate of N87 is about 0.13 against 0.49 in the classical circuit. The 32 parameters of the scheduled circuit therefore describe it only together with the network.

The energy exchanged by the time-varying elements was evaluated on 200 test windows per material in the 50/50 scenario (open-loop part). Per step, the energy that the schedule adds to the inductors at fixed state is compared with the energy dissipated in the sliders and relaxation resistors over the same window (the feedback sources are not counted; they are active in the classical circuit as well). Within a window, the schedule adds and removes energy of the same order as the dissipation, but on net it adds a median of 4 % of the dissipated energy (95th percentile 47 %), and only in 1.4 % of the windows more than is dissipated, most often on T37 (11 %) and 3E6 (9 %). The scheduled circuit is thus not passive in the strict sense, but it is dominated by dissipation in almost every window.

### C. Ablation

Table IV removes one classical element at a time from the classical circuit, with the same data and seed. Every element contributes. By far the most important is the measurement correction (10): without it the error rises by 5.3 percentage points and the GRU is ahead on all 15 materials; it is a property of the acquisition, but without it the model cannot follow the recorded $B$. Among the material elements, the mean-field feedback (+0.8, worse on 10 materials by more than 0.3 points) and the saturation knee (+0.6, up to +3.4 on N49) matter most, followed by the relaxations (+0.4) and the cubic term (+0.3). Rate-independent plays are as good on most materials but lose strongly on 3C90 and 77, where their fit fails. The lag of $H$ matters only for the lagged materials (3C92 +1.5, 3C95 +0.5).

Conversely, the model can be made smaller at a moderate cost (supplementary material): with a temperature-independent measurement correction and a common feedback gain for all operators, 27 parameters give 7.79 ± 0.08 % (ten seeds), and with three more coefficients independent of temperature, 24 parameters reach 7.81 %, about the accuracy of the GRU with a fourteenth of its parameters. Most of the loss falls on the materials with a measurement lag.

### D. Extrapolation and Data Efficiency

A model built from classical elements should need less data and degrade more gracefully outside its training range than a recurrent network. The classical circuit and a GRU8 replica (the RHINO-MAG architecture, 336 parameters, trained with the same pipeline) were fitted to four materials (3C94, N87, 3F4, 77) under five restrictions of the training data and evaluated on the test windows outside the restriction, next to the full-data model (Table V; the replica is weaker than the published network and serves only this comparison). Both need little data: trained on 2 % of the sequences, the model is as accurate as with all of them (6.5 % vs 6.4 %), and so is the GRU, at a higher level. Outside the training range they differ. Trained up to 320 kHz, the GRU fails at 500 and 800 kHz (54.5 %, up to 106 % on 77), while the model loses 4.9 points (10.2 % vs 5.3 %), most of it on N87. Across temperature, the GRU loses about twice as much: interpolating to 50 °C costs the model 0.8 points against 6.4, extrapolating to 70 °C 3.7 against 8.5, the latter mostly on 77.

TABLE V

Test SRE (%) on the windows outside the training restriction, restricted / full-data model, mean over 3C94, N87, 3F4 and 77 and the three scenarios. GRU8: the RHINO-MAG architecture trained with the same pipeline.

| Training data | Classical (32) | GRU8 (336) |
|---|---|---|
| 25, 50 °C → test at 70 °C | 9.3 / 5.6 | 16.4 / 7.9 |
| 25, 70 °C → test at 50 °C | 7.6 / 6.8 | 16.5 / 10.1 |
| $f \leq 320$ kHz → test at 500, 800 kHz | 10.2 / 5.3 | 54.5 / 6.0 |
| 10 % of the sequences | 6.5 / 6.4 | 9.4 / 9.6 |
| 2 % of the sequences | 6.5 / 6.4 | 9.1 / 9.6 |

## VIII. Discussion

**Why so few parameters suffice.** The GRU has to learn, from data alone, a representation of hysteresis memory. The play operators provide this memory structurally: their states store the reversal points of the input in exactly the way the wiping-out and congruency properties of Preisach-type models require [4]. The data then only have to determine a one-dimensional weight density over the operator widths, a small number of time constants and the reversible curve.

**Why gain scheduling works.** The classical circuit underfits: the constraints that make it readable cap its accuracy. Gain scheduling puts more parameters into the element values while the structure, the memory and the constraints of the circuit remain; what is learned is how the elements depend on the excitation, a dependence that classical hysteresis models introduce analytically, e.g., through rate-dependent densities [28]. Because the network sees only the excitation, the scheduled circuit remains a circuit driven by $B$ and $\vartheta$, and the schedule can be inspected: it acts at the reversals of $B$ and makes the hysteresis density depend on the amplitude. The price is that the 32 parameters lose their stand-alone meaning, since the circuit is described by them and the network together, and that the time-varying inductors make the circuit active in a small fraction of windows. Where readable parameters are needed, the classical circuit remains the model; where accuracy matters most, the scheduled circuit gives it at a small multiple of the parameters and without a recurrent network.

**Limitations.** The model stays behind the GRU on N87, 3C92 and FEC014 (Section VII.A). The warm-up estimator is learned per material. With only three training temperatures, the temperature dependence is restricted to linear terms, and DC-biased minor loops have not yet been tested.

## IX. Conclusion

For the MagNet Challenge 2 data, what a challenge-winning recurrent network has learned fits into a compact equivalent circuit of classical elements: a nonlinear inductor, a Prandtl–Ishlinskii bank of viscous play branches, two relaxation branches and a mean-field feedback, between a correction of the measured flux density and a lag of the measured field strength. With 32 parameters per material it is more accurate than the 336-parameter GRU (7.33 % vs 7.77 %) and better on 12 of 15 materials. It can be read: its parameters separate into the acquisition and the material, it revealed a lag in the measurement of the field strength, and its material part is valid at any time step and runs in circuit simulation. Gain-scheduled by small networks that set its element values from the excitation, it reaches 5.25 % with 656 parameters per material and 4.80 % with 5552, which would rank second of 33 in the challenge by both metrics.

The code (PyTorch with the fused Triton kernel), the fitted parameters of all materials with a NumPy implementation and SPICE subcircuits, the supplementary material, and the scripts that reproduce every table are available at https://github.com/wkirgsn/ferrite-bh-circuit. Open directions are an amplitude estimator for short warm-ups, a model-based observer without learned gains, a saturation description that extrapolates safely in temperature, schedules shared across materials or constrained to remain passive, and measurements of the same material with different acquisition chains.

## Acknowledgment

Claude Opus 5.5 (Anthropic) assisted in every facet of this work: the literature research, the analysis of the data, the design, implementation and training of the models and of the GPU kernel, the experiments and their evaluation, the equivalent-circuit verification, the figures, and the writing of this manuscript.